\documentclass{PoS} \PoS{PoS(LAT2005)293}
\FullConference{XXIIIrd International Symposium on Lattice Field Theory\\
         25-30 July 2005\\
         Trinity College, Dublin, Ireland}

\usepackage{graphicx}

\title{Gribov horizon under the (lattice) microscope%
\addtocounter{footnote}{1}\thanks{Our research is supported in part
by the U.S. Department of Energy under Grant No.\ DE-FG03-92ER40711
(J.G.), the Slovak Science and Technology Assistance Agency, Grant
No.\ APVT--51--005704 (\v{S}.O.), and the National Science
Foundation, Grant No.\ PHY-0099393 (D.Z.).}}
\ShortTitle{Gribov horizon under the (lattice) microscope}
\author{Jeff Greensite\\
        The Niels Bohr Institute, Blegdamsvej 17, DK-2100 Copenhagen \O, Denmark;\\
        Physics and Astronomy Dept., San Francisco State University,
        San Francisco, CA~94117, USA\\
        E-mail: \email{greensit@stars.sfsu.edu}}
\author{\addtocounter{footnote}{-2}\speaker{\v Stefan Olejn\'\i k}\\
        Institute of Physics, Slovak Academy of Sciences,
        SK--845 11 Bratislava, Slovakia\\
        E-mail: \email{stefan.olejnik@savba.sk}}
\author{Daniel Zwanziger\\
        Physics Department, New York University, New York, NY~10003, USA\\
        E-mail: \email{daniel.zwanziger@nyu.edu}}

\abstract{The infinite color-Coulomb energy of color-charged states
is related to enhanced density of near-zero modes of the
Faddeev--Popov operator in Coulomb gauge. We confirm the enhancement
in numerical simulations and show that it is tied to the presence of
percolating center vortex configurations.}

\begin{document}


\section{Introduction: Confinement scenario in Coulomb gauge}

    The Faddeev--Popov operator in Coulomb gauge
\begin{equation}\label{FPOperator}
M(A)\equiv-\nabla\cdot{\cal{D}}(A), \qquad
\mbox{where}\qquad{\cal{D}}^{ac}_i(A)=\partial_i\delta^{ac}+f^{abc}A^b_i(x),
\end{equation}
plays a crucial role in the Gribov--Zwanziger confinement
scenario~\cite{Gribov:1977wm,Zwanziger:1998ez}. It enters the kernel
$K$ in the (classical) Coulomb energy of a color charge distribution
$\rho$:
\begin{equation}\label{hamiltonian}
H_{coul}=\frac{1}{2}\int d^3\mathbf{x}d^3\mathbf{y}\; \varrho^a(x)
K^{ab}(x,y;A)\varrho^b(y),\qquad \varrho^a=\varrho^a_{matter} -
f^{abc}\mathbf{A}^b\cdot\mathbf{E}^c,
\end{equation}
where
\begin{equation}
 K^{ab}(x,y;A)\equiv\left[M^{-1}\;(-\nabla^2)\;M^{-1}\right]^{a,b}_{x,y}.
\end{equation}

    The essence of the scenario can be summarized in the following
way: The Coulomb-gauge condition $\nabla\cdot\mathbf{A}^a=0$ does
not fix the gauge completely. Gribov~\cite{Gribov:1977wm} suggested
to restrict to the subspace of transverse gauge fields for which the
Faddeev--Popov operator is positive, i.e.\ local minima with respect
to $g(x)$ of
\begin{equation}\label{functional}
I[\mathbf{A},g]=\int dx\left[^g \mathbf{A}^a(x)\right]^2,\quad
\mbox{where}\quad{}^gA_i=g^{-1} A_i g +g^{-1}\partial_i g.
\end{equation}
The boundary of this Gribov region (GR) is called the \emph{Gribov
horizon}. However, even this does not eliminate the Coulomb-gauge
ambiguities completely, one has to further narrow the gauge-field
configuration space to the \emph{fundamental modular region} (FMR),
i.e.\ the set of absolute minima of the functional
(\ref{functional}). Both the GR and the FMR are bounded in every
direction and convex. The dimension of the gauge-field configuration
space is huge, so it is reasonable to expect that most
configurations are located close to its boundary (horizon). The
interaction kernel $K$ contains the inverse of the FP operator,
which is strictly zero on the horizon and near-zero close to the
horizon. A high density of configurations near the horizon leads to
a strong enhancement of the Coulomb interaction energy, and
hopefully causes color confinement.

    In this contribution we formulate a simple criterion of confinement for
static color charges through properties of eigenstates of the FP
operator in Coulomb gauge close to the Gribov horizon, and then
discuss how the fulfillment of this criterion depends on
presence/absence of center vortices. Details, as well as some
analytic insights on the connections between center vortices and the
Gribov horizon, can be found in a recent
publication~\cite{Greensite:2004ur}.\footnote{We have also
investigated localization properties of the lowest nontrivial
eigenvectors of the Faddeev--Popov operator in Coulomb gauge. These
were discussed in Jeff Greensite's talk at this conference and in
Sect.\ V of Ref.\ \cite{Greensite:2005ur}.}

\section{Lattice Faddeev--Popov operator and its eigenstates}

    If we parametrize link variables in SU(2)
lattice gauge theory by
\begin{equation}\label{U}
U_\mu(x)=b_\mu(x)+i\sigma^c a_\mu^c(x), \qquad b_\mu(x)^2+\sum_c
a_\mu^c(x)^2=1,
\end{equation}
the lattice Faddeev--Popov operator in Coulomb gauge is given by the
following expression:
\begin{eqnarray}\label{latticeFP}
\nonumber
  M^{ab}_{xy} &=& \delta^{ab} \sum_{k}\left\{ \delta_{xy}
\left[b_k(x)
   + b_k(x-\hat{k})\right] - \delta_{x,y-\hat{k}} b_k(x)
   -  \delta_{y,x-\hat{k}} b_k(y) \right\}\\
   &-& \epsilon^{abc} \sum_{k} \left\{ \delta_{x,y-\hat{k}}
a^c_k(x) - \delta_{y,x-\hat{k}} a^c_k(y)  \right\}.
\end{eqnarray}
We are interested in its eigenstates
\begin{equation}\label{eigenequation}
{\sum_{b,y}}M^{ab}_{xy}\;\phi^{(n)b}_{y}=\lambda_n\;\phi^{(n)a}_{x},
\end{equation}
in particular in their properties near to the Gribov horizon (i.e.\
in the limit $\lambda\to 0$). The most relevant quantities are
\begin{list}{$\bullet$}{\topsep3pt\parsep0pt\itemsep3pt}
\item the density of eigenstates $\rho(\lambda)$, and
\item the average Laplacian:
\begin{equation}
F_{n}\equiv{\displaystyle{\sum_{a,xy} \phi^{(n)a}_{x}
(-\nabla^2)_{xy}\phi^{(n)a\ast}_{y}}}.
\end{equation}
\end{list}

\section{A confinement condition}

    We shall now formulate a simple confinement criterion in terms
of properties of eigenstates of the FP operator. The energy of a
static color charge state $\Psi_C^\alpha[A;x]$ in Coulomb gauge
\begin{equation}\label{staticcharge}
{\cal{E}} =\frac{\langle \Psi^\alpha_C \vert H_{coul} \vert
\Psi^\alpha_C \rangle} {\langle \Psi^\alpha_C \vert \Psi^\alpha_C
\rangle} - \langle \Psi_0 \vert H_{coul} \vert \Psi_0 \rangle\sim
\langle K^{aa}(x,x;A) \rangle
\end{equation}
can be easily shown to be given by
\begin{equation}\label{Ethrueigenstates}
{{\cal{E}}=\frac{1}{3V_3}\sum_n
\left\langle\frac{F_n}{\lambda_n^2}\right\rangle}\qquad\mbox{going
to}\qquad
\int_0^{\lambda_{max}}\frac{d\lambda}{\lambda^2}\left\langle\rho(\lambda)F(\lambda)\right\rangle
\quad\mbox{ for }V_3\rightarrow\infty.
\end{equation}

    An immediate consequence is that
\emph{the excitation energy ${\cal{E}}$ of a static, unscreened
color charge is divergent if, at infinite volume,}
\begin{equation}\label{condition}
\lim_{\lambda\to0}\frac{\left\langle\rho(\lambda)F(\lambda)\right\rangle}{\lambda}>0.
\end{equation}
This criterion is a necessary but not sufficient condition for
confinement; an explicit example will be given at the end of
Sect.~\ref{results}. (It is obviously not fulfilled in the free
theory, where $\rho(\lambda)\sim\sqrt{\lambda}$,
$F(\lambda)=\lambda$, and consequently
${\cal{E}}\sim\sqrt{\lambda_{max}}$.)

\section{Three ensembles of lattice configurations}

    We will investigate fulfillment of the
condition~(\ref{condition}) in three ensembles of configurations:
\begin{list}{}{\topsep3pt\parsep0pt\itemsep3pt}
\item[1.] \textbf{\emph{full configurations}},
    $\lbrace U_\mu(x)\rbrace$;
\item[2.] \textbf{\emph{``vortex-only'' configurations}}: these are
    obtained from full configurations fixed to the (direct) maximal center
    gauge~\cite{DelDebbio:1998uu} by center projection, $\lbrace Z_\mu(x) = \mbox{sign
    Tr}[U^{(MCG)}_\mu(x)]\rbrace$;
\item[3.] \textbf{\emph{``vortex-removed'' configurations}}, obtained by the
    recipe of de Forcrand and D'Elia \cite{deForcrand:1999ms}:
    $\lbrace U_\mu^{(R)}(x) = Z_\mu^\dagger(x) U_\mu(x)\rbrace$.
\end{list}

    Each configuration in these three ensembles was brought to Coulomb
gauge by maximizing with respect to gauge transformations, on each
time slice ${\cal{R}}_{coul}(t)={\sum_{\mathbf{x}}\sum_{k=1}^3}
{\textstyle{\frac{1}{2}}}\mbox{Tr}[U_k(\mathbf{x},t)]. $
%

\section{Results}\label{results}

\begin{figure}[t!]
\begin{center}
\begin{tabular}{c p{0.01\hsize} c}
  \includegraphics[width=0.43\hsize]{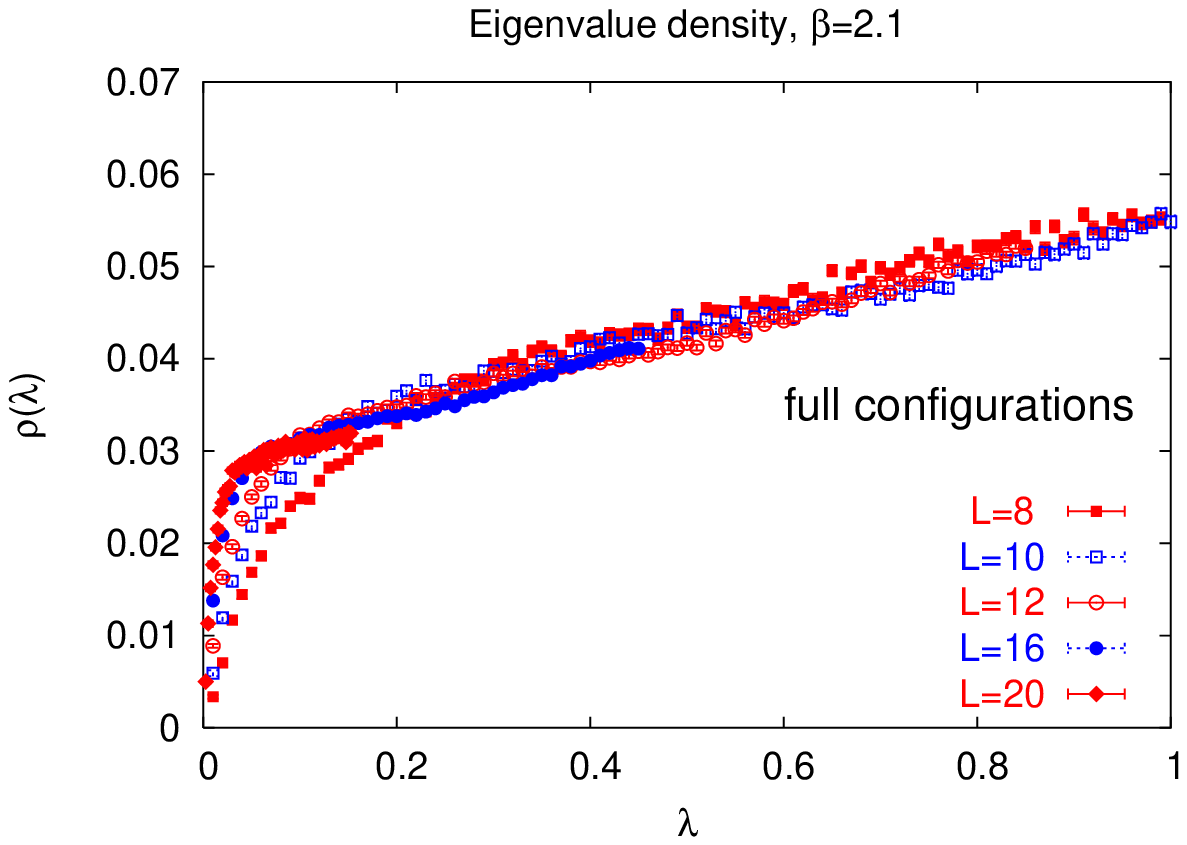}& &
  \includegraphics[width=0.43\hsize]{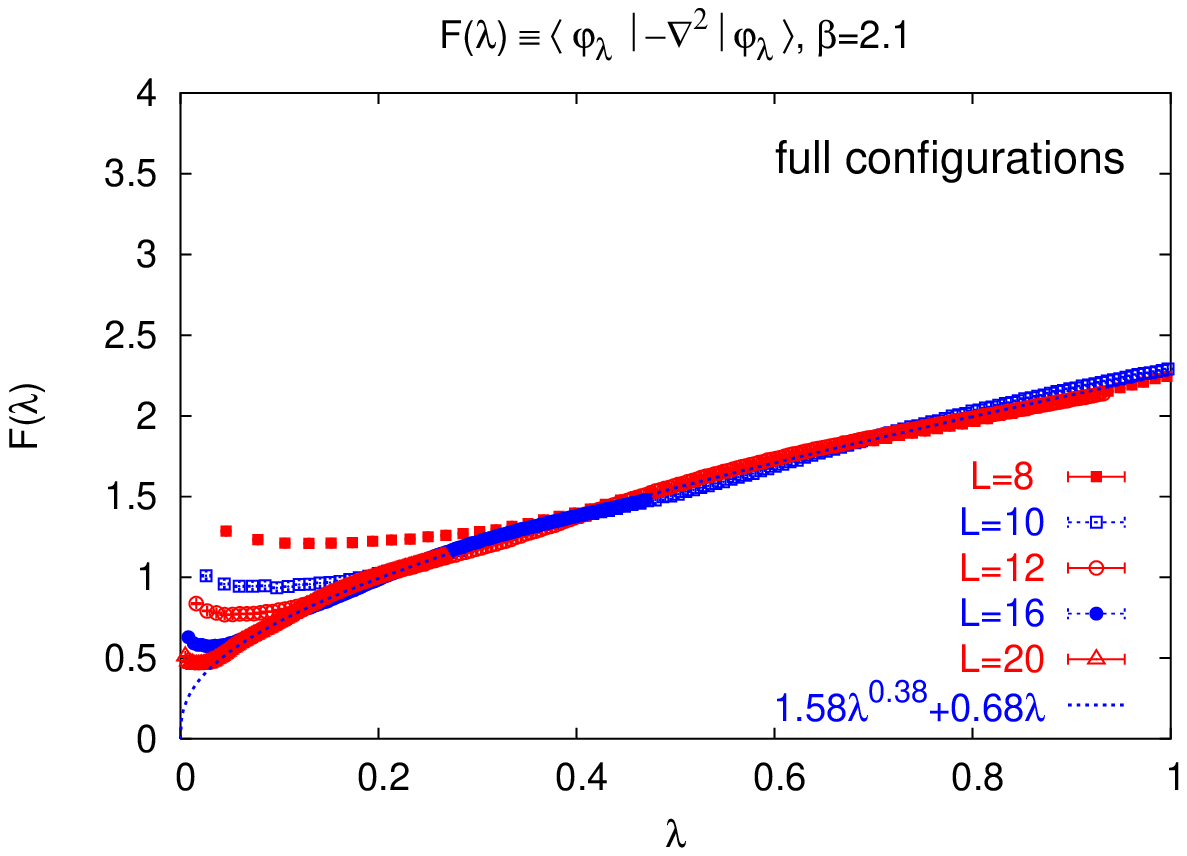}\\
\end{tabular}
\end{center}
\caption{$\rho(\lambda)$ and $F(\lambda)$ for full lattice
configurations.} \label{2p1_all_full}
\end{figure}
    \textbf{\emph{Pure gauge theory at zero temperature.}} The
results for \emph{full configurations} at $\beta=2.1$ are shown in
Figure~\ref{2p1_all_full}, for a series of lattice volumes.%
\footnote{The results for $\beta=2.3$ and $2.4$ can be found in
Ref.~\cite{Greensite:2004ur}, and are qualitatively the same as
those for $\beta=2.1$.} Both $\rho(\lambda)$ and $F(\lambda)$
exhibit a sharp ``bend'' near $\lambda\to 0$, and behave near $0$
like a small power of $\lambda$. A scaling analysis similar to that
used in random matrix theory gives the estimates
\begin{equation}\label{powersfull}
\rho(\lambda)\sim\lambda^{0.25},\qquad F(\lambda)\sim\lambda^{0.38}.
\end{equation}
The confinement condition (\ref{condition}) is obviously satisfied,
which is a direct manifestation of the mechanism proposed by Gribov
and Zwanziger.

\begin{figure}[b!]
\begin{center}
\begin{tabular}{c p{0.01\hsize} c}
  \includegraphics[width=0.43\hsize]{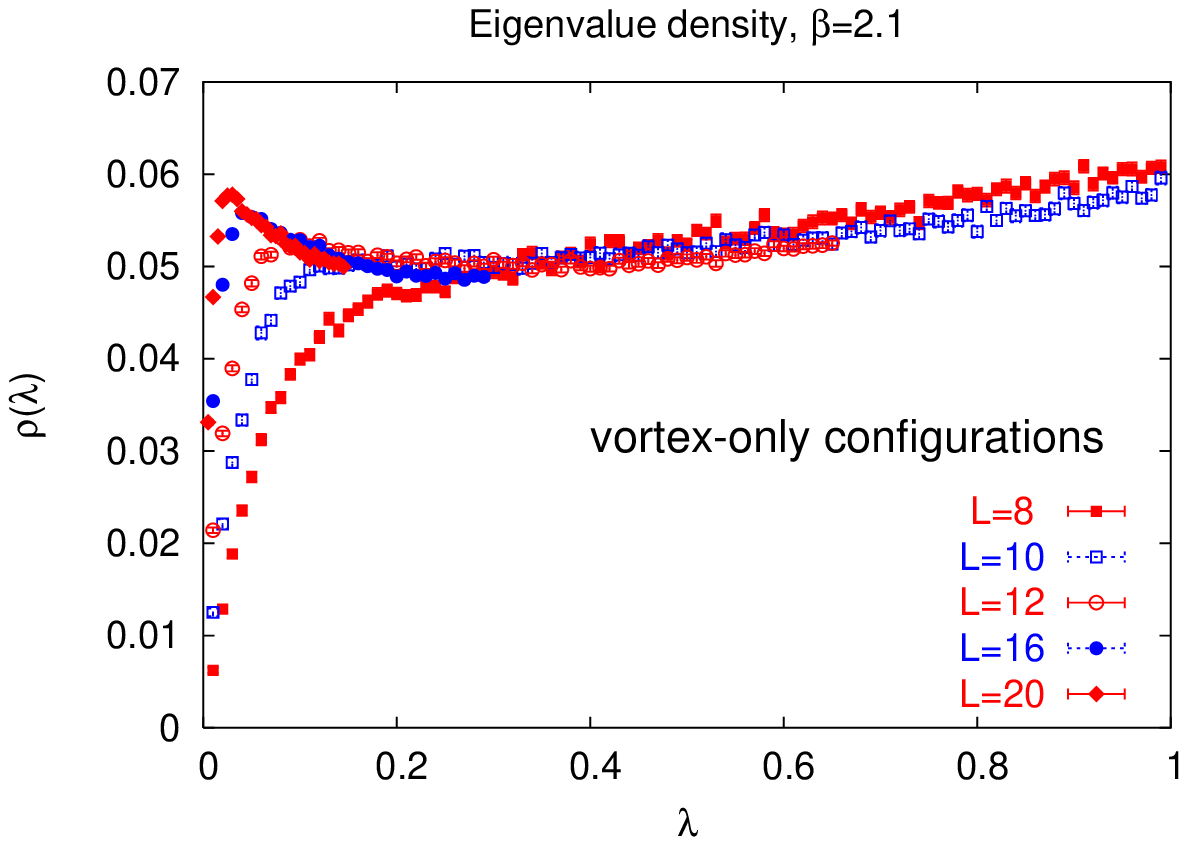}& &
  \includegraphics[width=0.43\hsize]{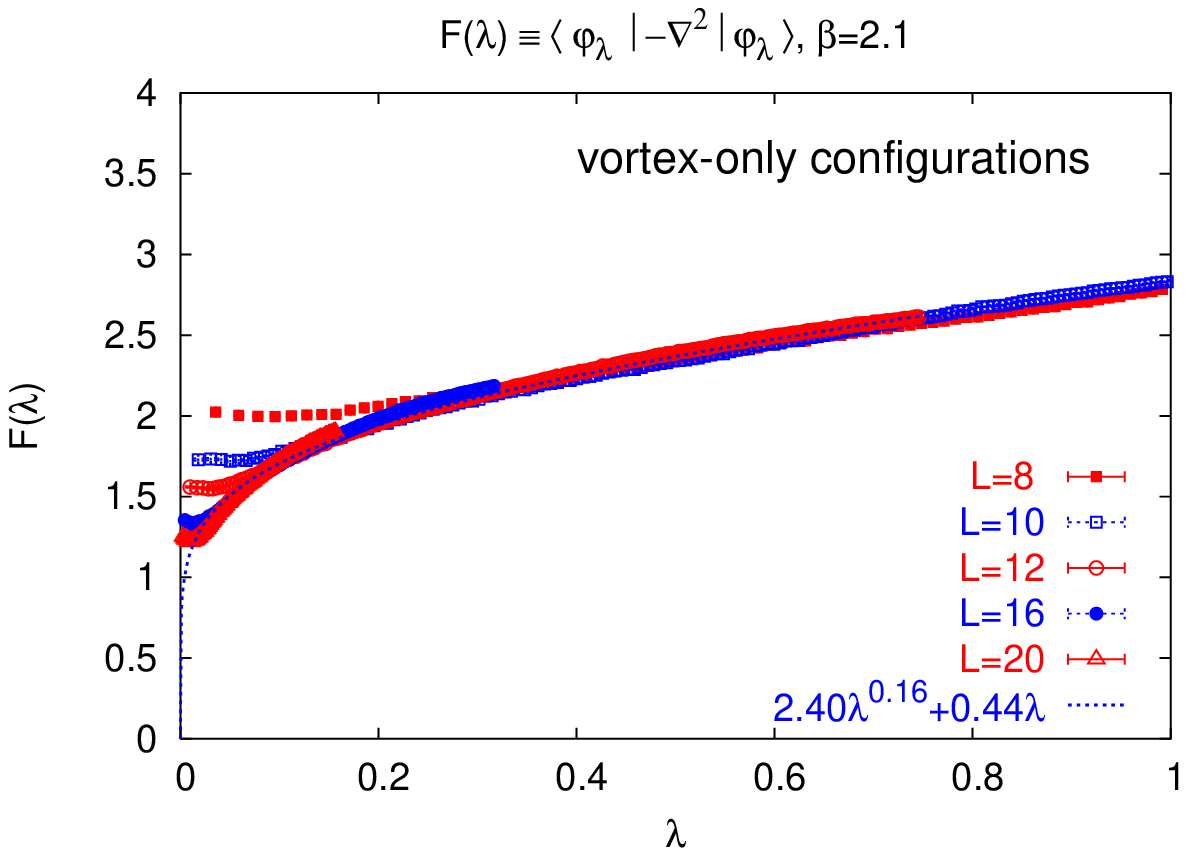}\\
\end{tabular}
\end{center}
\caption{$\rho(\lambda)$ and $F(\lambda)$ for vortex-only
(center-projected) configurations.} \label{2p1_all_cp}
\end{figure}
    The situation in \emph{vortex-only configurations} is displayed
in Figure~\ref{2p1_all_cp}. The enhancement of the density of states
is even more pronounced than in full configurations, and both
quantities of interest seem to converge to a non-zero value in the
infinite volume limit
\begin{equation}\label{powerscp}
\rho(0)\sim 0.06,\qquad F(0)\sim 1.0.
\end{equation}
(though their proportionality to very small powers of $\lambda$
cannot be excluded). Once again, the condition~(\ref{condition}) is
fulfilled.

\begin{figure}[t!]
\begin{center}
\begin{tabular}{c p{0.01\hsize} c}
  \includegraphics[width=0.43\hsize]{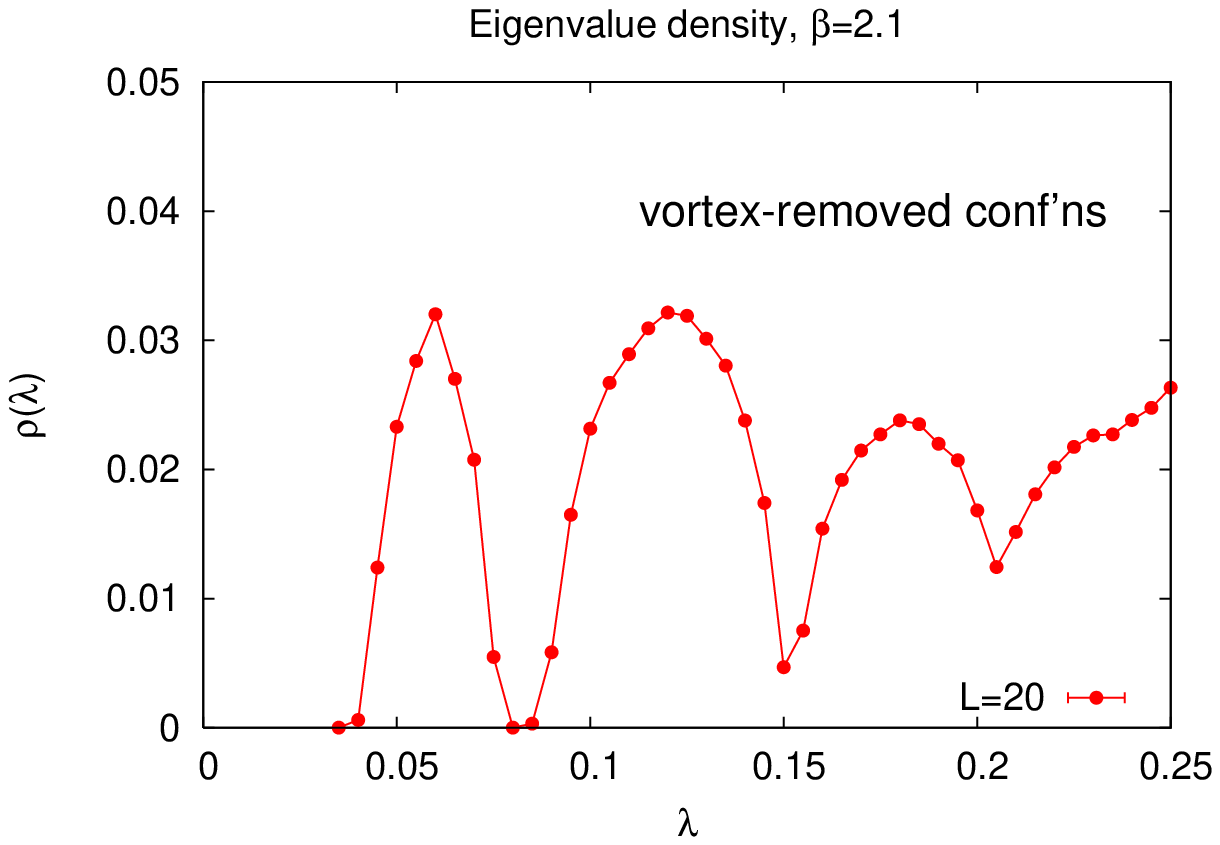}& &
  \includegraphics[width=0.43\hsize]{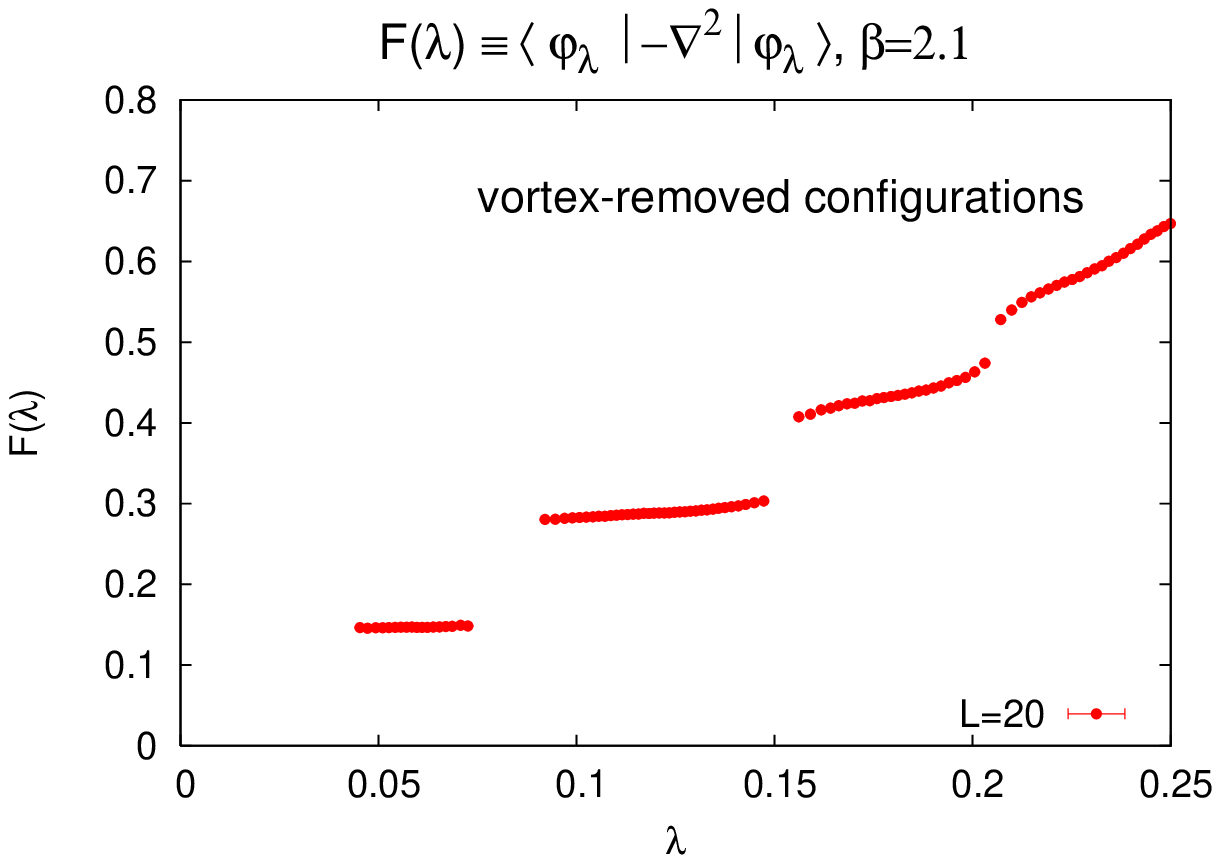}\\
\end{tabular}
\end{center}
\caption{$\rho(\lambda)$ and $F(\lambda)$ for vortex-removed
configurations.} \label{2p1_nv}
\end{figure}
    The eigenvalue spectrum of the FP operator is drastically
different for \emph{vortex-removed configurations}, see
Figure~\ref{2p1_nv} for the largest available, $20^4$ lattice. The
density displays a series of peaks, and values of $F(\lambda)$ are
organized into bands, separated by gaps. This can be understood
rather simply: For the Laplacian operator (equal to the FP operator
at zero-th order in the gauge coupling) the eigenvalue density, at
finite volume, is a sum of delta-functions, and each eigenvalue is
multiply degenerate. The vortex-removed configuration seems to be
just a small perturbation around the zero-coupling limit, which
lifts the degeneracy. In this way, delta-functions in the density of
states turn into distinct peaks of finite width, and degenerate
values of $F(\lambda)$ spread into bands. The number of values
inside the $k$-th band of $F(\lambda)$ exactly matches the
degeneracy of the $k$-th eigenvalue of the unperturbed Laplacian
operator.

    This result demonstrates a deep relation between the
Gribov-horizon and center-vortex confinement mechanism. Center
vortices seem to be \emph{the} field configurations providing the
mechanism needed for enhancement of eigenvalues near the horizon.

\begin{figure}[b!]
\begin{center}
\begin{tabular}{c p{0.01\hsize} c}
  \includegraphics[width=0.43\hsize]{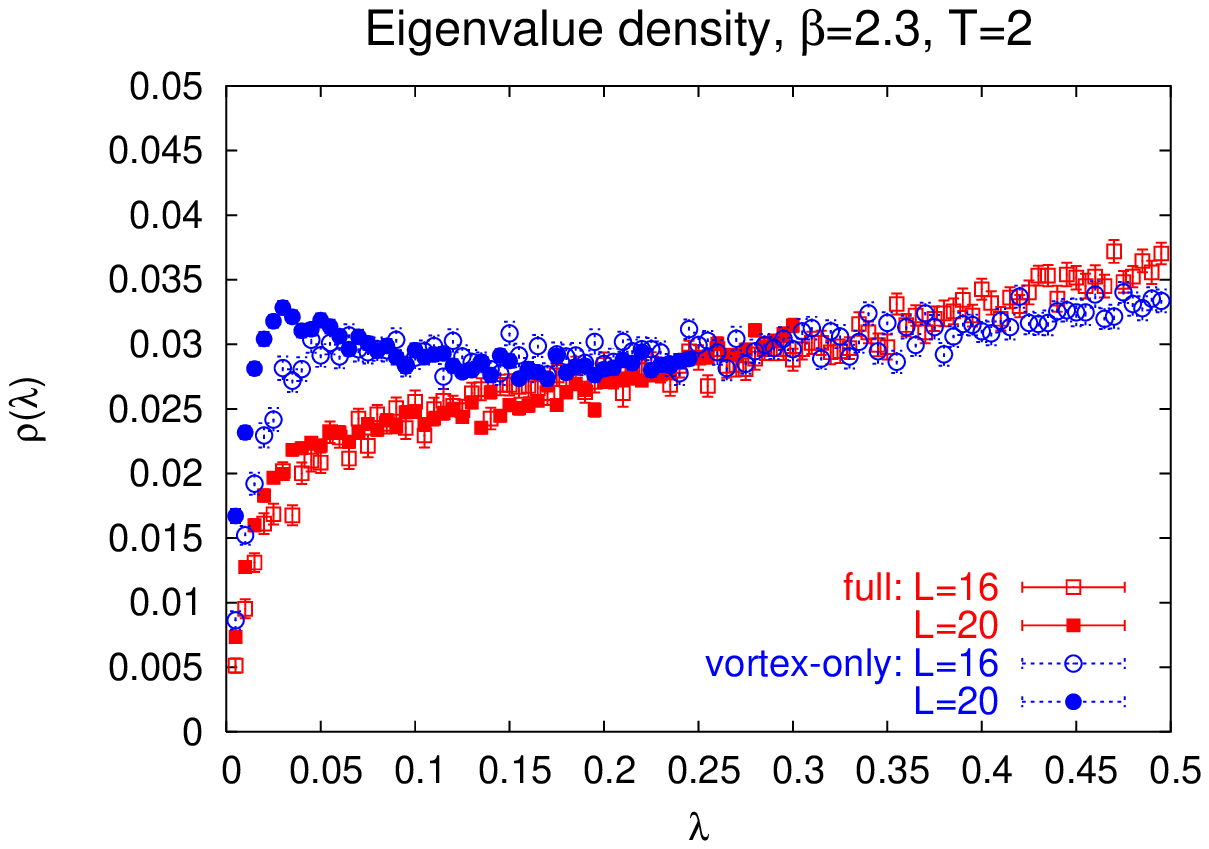}& &
  \includegraphics[width=0.43\hsize]{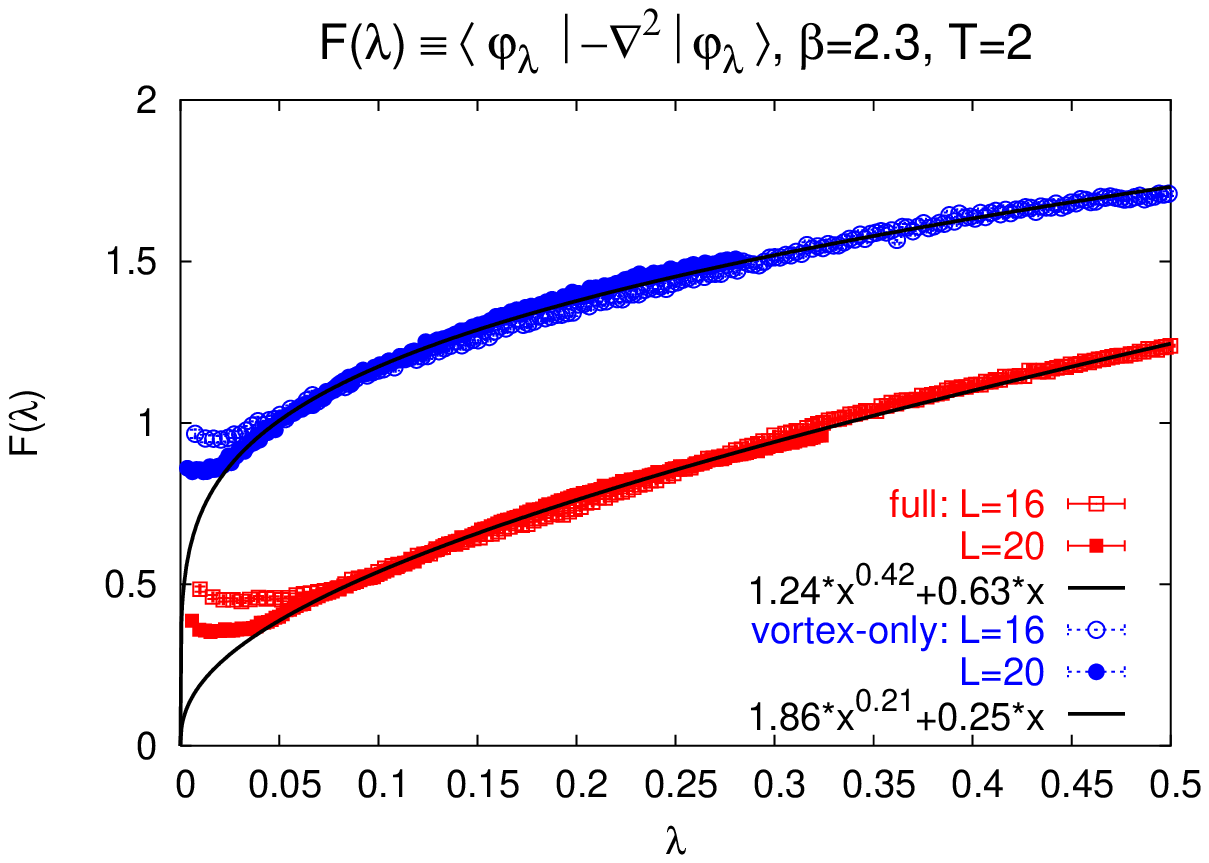}\\
\end{tabular}
\end{center}
\caption{$\rho(\lambda)$ and $F(\lambda)$ in the deconfined phase,
for full and vortex-removed configurations.} \label{2p3_all_t2}
\end{figure}
    \textbf{\emph{Pure gauge theory in the deconfined phase.}} A
seemingly paradoxical result is obtained above the deconfinement
transition: our quantities above $T_c$ look the same as at $T=0$
(\emph{cf.}\ Figure~\ref{2p3_all_t2} with Figs.~\ref{2p1_all_full}
and \ref{2p1_all_cp})! However, one should keep in mind that
spacelike links are a confining ensemble even in the deconfined
phase, and spacelike Wilson loops have an area law behavior.

    The result for the deconfined phase can be naturally explained in the
Gribov-horizon scenario. In Coulomb gauge the gauge fixing is done
independently on each 3$d$ time slice. According to the horizon
scenario, on each time slice, 3$d$ configurations $\mathbf{A}({\bf
x})$ are favored that lie near the horizon of a 3$d$ gauge theory,
and this enhances the instantaneous color-Coulomb potential. This is
true for {\it every} temperature $T$, including in the deconfined
phase, because temperature determines the extent of the lattice in
the {\it fourth} dimension. Thus, the horizon scenario provides a
framework in which confinement may be understood, but it is not
detailed enough to tell us under what conditions the infinite
color-Coulomb potential may be {\it screened} to give a finite
self-energy.\footnote{For another example of operation of the
horizon scenario in the deconfined phase see Ref.\
\cite{Zwanziger:2004np}.}


\section{Conclusions}

    The low-lying eigenvalues of the FP operator in Coulomb gauge tend
towards zero as the lattice volume increases. The density of the
eigenvalues goes as a small power of $\lambda$, and this, together
with a similar behavior of the average Laplacian, $F(\lambda)$,
assures the infrared divergence of the energy of an unscreened color
charge. These facts \textit{support the ideas of the Gribov-horizon
confinement scenario}.

    The constant density of low-lying eigenvalues can be attributed to
the vortex component of gauge-field configurations. A thermalized
configuration in a pure gauge theory factors into a confining piece
(the vortex-only part), and a piece which closely resembles the
lattice of a gauge--Higgs theory in the Higgs phase (the
vortex-removed configuration). This establishes \textit{firm
connection between the center-vortex picture and the Gribov-horizon
scenario}.

    The Gribov--Zwanziger scenario, though invented to explain
confinement, is operative also in the finite temperature deconfined
phase.

    Here we only covered results of our numerical investigations. Related
analytical developments were omitted
and can be found in our
recent publication~\cite{Greensite:2004ur}.


\end{document}